\documentclass[superscriptaddress,preprint,aps]{revtex4}%
\usepackage{amsfonts}
\usepackage{amsmath}
\usepackage{amssymb}
\usepackage{graphicx}
\usepackage{xcolor}%
\providecommand{\U}[1]{\protect\rule{.1in}{.1in}}

\begin{document}
\title{The soft-mode phonons mediated unconventional superconductivity in monolayer
1T$^{\prime}$-WTe$_{2}$}
\author{Wei Yang}
\affiliation{Beijing Key Laboratory of Work Safety Intelligent Monitoring, Beijing
University of Posts and Telecommunications, Beijing 100876, China}
\author{Chong-jie Mo}
\affiliation{Beijing Computational Science Research Center, Beijing 100193, China}
\author{Shi-Bin Fu}
\affiliation{Beijing Key Laboratory of Work Safety Intelligent Monitoring, Beijing
University of Posts and Telecommunications, Beijing 100876, China}
\author{Yu Yang}
\affiliation{Institute of Applied Physics and Computational Mathematics, Beijing 100088, China}
\author{Fa-Wei Zheng}
\affiliation{Institute of Applied Physics and Computational Mathematics, Beijing 100088, China}
\author{Xiao-Hui Wang}
\affiliation{Beijing Key Laboratory of Work Safety Intelligent Monitoring, Beijing
University of Posts and Telecommunications, Beijing 100876, China}
\author{Yuan-An Liu}
\affiliation{Beijing Key Laboratory of Work Safety Intelligent Monitoring, Beijing
University of Posts and Telecommunications, Beijing 100876, China}
\author{Ning Hao}
\email{haon@hmfl.ac.cn}
\affiliation{Anhui Province Key Laboratory of Condensed Matter Physics at Extreme
Conditions, High Magnetic Field Laboratory, HFIPS,Chinese Academy of Sciences,
Hefei, 230031, China}
\author{Ping Zhang}
\email{zhang_ping@iapcm.ac.cn}
\affiliation{Institute of Applied Physics and Computational Mathematics, Beijing 100088, China}
\affiliation{School of Physics and Physical Engineering, Qufu Normal University, Qufu
273165, China}
\affiliation{Beijing Computational Science Research Center, Beijing 100193, China}

\begin{abstract}
Recent experiments have tuned the monolayer 1T$^{\prime}$-WTe$_{2}$ to be
superconducting by electrostatic gating. Here, we theoretically study the
phonon-mediated superconductivity in monolayer 1T$^{\prime}$-WTe$_{2}$ via
charge doping. We reveal that the emergence of soft-mode phonons with specific
momentum is crucial to give rise to the superconductivity in electron-doping
regime, whereas no such soft-mode phonons and no superconductivity emerge in
hole-doping regime. We also find a superconducting dome, which can be
attributed to the change of Fermi surface nesting condition as electron
doping. By taking into account the experimentally established strong
anisotropy of temperature-dependent upper critical field $H_{c2}$ between the
in-plane and out-of-plane directions, we show that the superconducting state
probably has the unconventional equal-spin-triplet pairing in $A_{u}$ channel
of $C_{2h}$ point group. Our studies provide a promising understanding to the
doping dependent superconductivity and strong anisotropy of $H_{c2}$ in
monolayer 1T$^{\prime}$-WTe$_{2}$, and can be extended to understand the
superconductivity in other gated transition metal dichalcogenides.

\end{abstract}
\maketitle

Tuning the topological materials to be superconducting provides a highly
efficient way to search and study the exotic superconductivity such as
unconventional and topological superconductivity. Some methods have been
developed to achieve the target, including the doping through metal
intercalation \cite{hor-prl-2010,liu-jacs-2015,smylie-prb-2017,asaba-prx-2017}%
, high pressure \cite{kirshenbaum-prl-2013,zhang-prb-2011}, proximity effect
\cite{fu-prl-2008,Lutchyn-prl-2010,xu-prl-2014,hao-nsr-2019}, hard and soft
tip contact
\cite{wang-sb-2018,wang-nm-2016,aggarwal-nm-2016,hou-prb-2019,hou-prb-2020},
and electrostatic gating \cite{Sajadi-s-2018,fatemi-s-2018}. Among them, the
electrostatic gating has advantage to freely tune the materials in both
electron- and hole-doped regimes without introducing dopant. Recently, the
intrinsic superconductivity in the monolayer topological insulator
1T$^{\prime}$-WTe$_{2}$ have been experimentally observed by two groups
through moderate electrostatic gating \cite{Sajadi-s-2018,fatemi-s-2018}. It
is found that the superconductivity shows some interesting features. For
instance, only the electron-doped regime show the superconductivity with the
transition temperature ($T_{c}$) up to 1 K, whereas no superconducting
signature is observed in the hole-doped regime. Furthermore, the upper
critical field $H_{c2}$ shows strong anisotropy between the in-plane
$H_{c2,\parallel}$ and out-of-plane $H_{c2,\perp}$, and $H_{c2,\parallel}$ is
significantly larger and four times the Pauli paramagnetic limit 1.84$T_{c}$.
However, the gating-dependent superconductivity and the $H_{c2}$ anisotropy
have not been comprehensively understood.

In this work, we show that the phonon spectrum of 1T$^{\prime}$-WTe$_{2}$
dramatically softens in the electron-doped regime, but slightly stiffens in
the hole-doped regime through density functional calculations. According to
the phonon-mediated superconductivity theory, we reveal that the softening of
the phonons with specific momentum in electron-doped regime is the driving
force to give rise to the superconductivity. We also find an optimal
electron-doped concentration, beyond which, the $T_{c}$ becomes to decline in
accompany with the latent charge density wave (CDW) instability. Furthermore,
we reveal the evolution of superconductivity coincides with the change of
Fermi surface nesting condition, and predict the CDW order is quasi-one
dimensional commensurate with wave vector $q_{CDW}=$(0, $\pi/b$) according to
the electronic susceptibility. Thus, there exist a superconducting dome and a
new CDW state in the phase diagram. Based on the symmetry classification and
linearized gap equations, we find that all the superconducting pairing
channels with specific irreducible representations (IRs) are degenerate
without external magnetic field. However, only the pairing in $A_{u}$ channel
shows the experimentally observed reasonable anisotropy between
$H_{c2,\parallel}$ and $H_{c2,\perp}$. This behavior indicates that the
superconductivity in 1T$^{\prime}$-WTe$_{2}$ probably has equal-spin-triplet
pairing and belongs to unconventional type.

The first-principles calculations in this work are performed using the density
functional theory (DFT) and the density functional perturbation theory
(DFPT)\cite{Perdew-prb-1981,Ceperley-prl-1980}. We employed the QUANTUM
ESPRESSO (QE) package\cite{Giannozzi-jpcm-2009} for the electronic structure
and lattice dynamics including the phonon spectrum and electron-phonon
coupling. The superconducting transition temperature $T_{c}$ is calculated
with the McMillan-Allen-Dynes formula\cite{allen-prb-1975} as implemented in
the QE package\cite{Giannozzi-jpcm-2009}. The details of the calculation are
presented in the supplementary materials (SM)\cite{sms}.

\begin{figure}[ptb]
\begin{center}
\includegraphics[width=1\columnwidth]{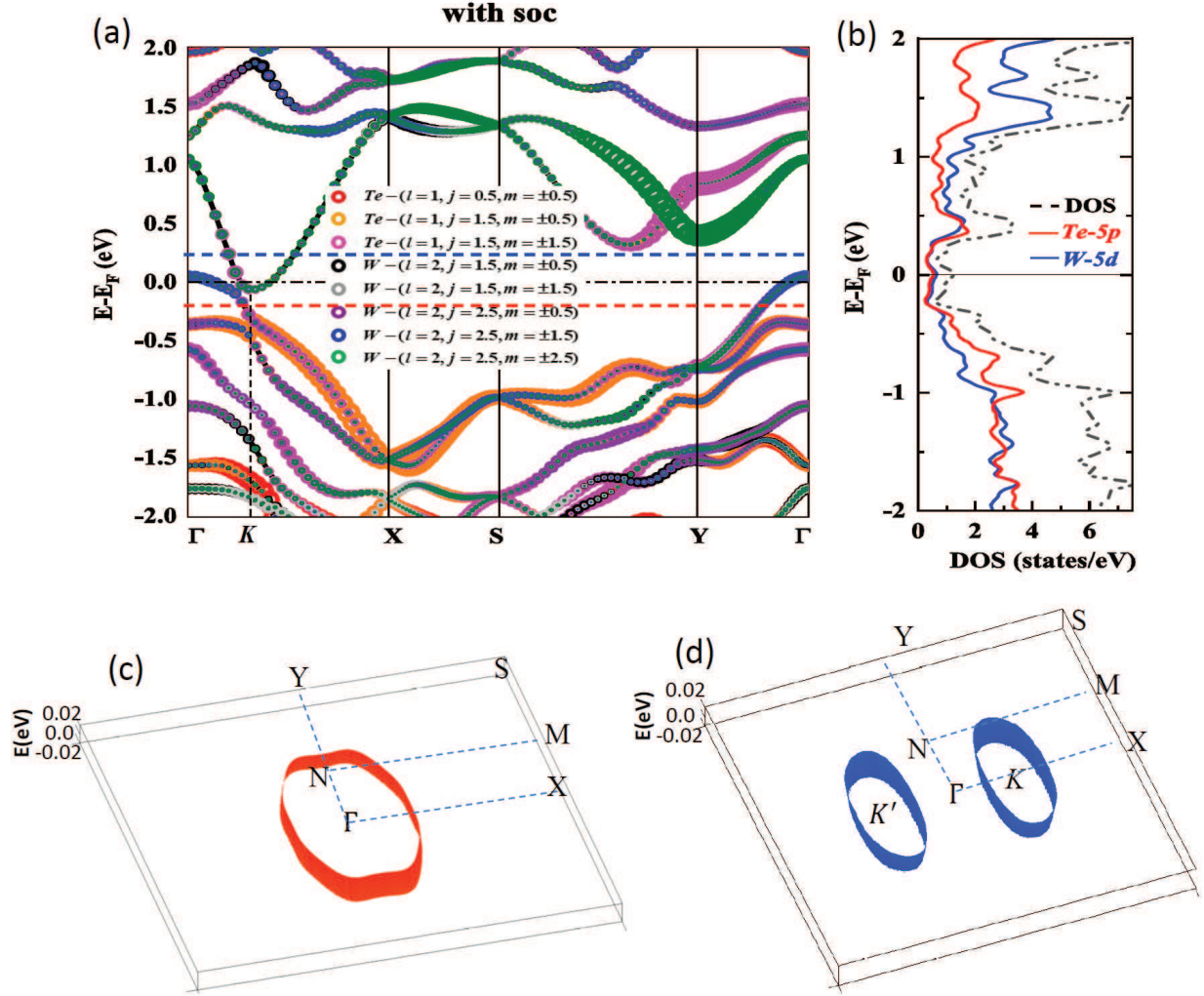}
\end{center}
\caption{(Color online)(a) The calculated orbital-resolved band structures of
1T$^{\prime}$-WTe along the high-symmetry lines with spin-orbit coupling. The
weight of the orbital is labeled by the linewidth. The conduction band bottom
is labeled by $K=(0.17,0)$ in units $\frac{2\pi}{a}$ and $\frac{2\pi}{b}$
($a=3.49$\AA $\ $and $b=6.31$\AA \ are the lattice constants
\cite{yang-prb-2019}). (b) The density of states of band structures in (a).
(c) and (d) the Fermi surfaces for the hole doping with $n_{h}=-9.5\times
10^{13}$ cm$^{-2}$ and electron doping with $n_{e}=9.5\times10^{13}$cm$^{-2}$,
respectively. The plotting is set with energy window [-0.02eV, 0.02eV]. The
Fermi levels with 0eV correspond to the red-dashed and blue-dashed lines in
(a), respectively.}%
\label{f1}%
\end{figure}

The charge doping effect from the electrostatic gating is simulated by adding
or removing electrons to the monolayer 1T$^{\prime}$-WTe$_{2}$ with a
compensating uniform charge background. The doped carrier concentrations per
1T$^{\prime}$-WTe$_{2}$ is expressed as $n$ (cm$^{-2}$), with positive and
negative values indicating electron and hole doping, respectively. Figure
\ref{f1} shows the electronic structure of 1T$^{\prime}$-WTe$_{2}$ from the
first-principle calculations. The Fermi surfaces in Figs. \ref{f1} (c) and (d)
correspond to the hole and electron doping with carrier concentrations
$n_{h}=-9.5\times10^{13}$ cm$^{-2}$ and $n_{e}=9.5\times10^{13}$cm$^{-2}$,
respectively. Around the Fermi level, the $d$ orbitals of W and $p$ orbitals
of Te are dominated. Along $\Gamma-X$ line, there are two electron pockets
labeled by $K^{\prime}$ and $K$ in electron-doped case while only one hole
pocket centered at $\Gamma$ point in hole-doped case. The details of
calculations are presented in Section I and II in SM. 

\begin{figure}[ptb]
\begin{center}
\includegraphics[width=1\columnwidth]{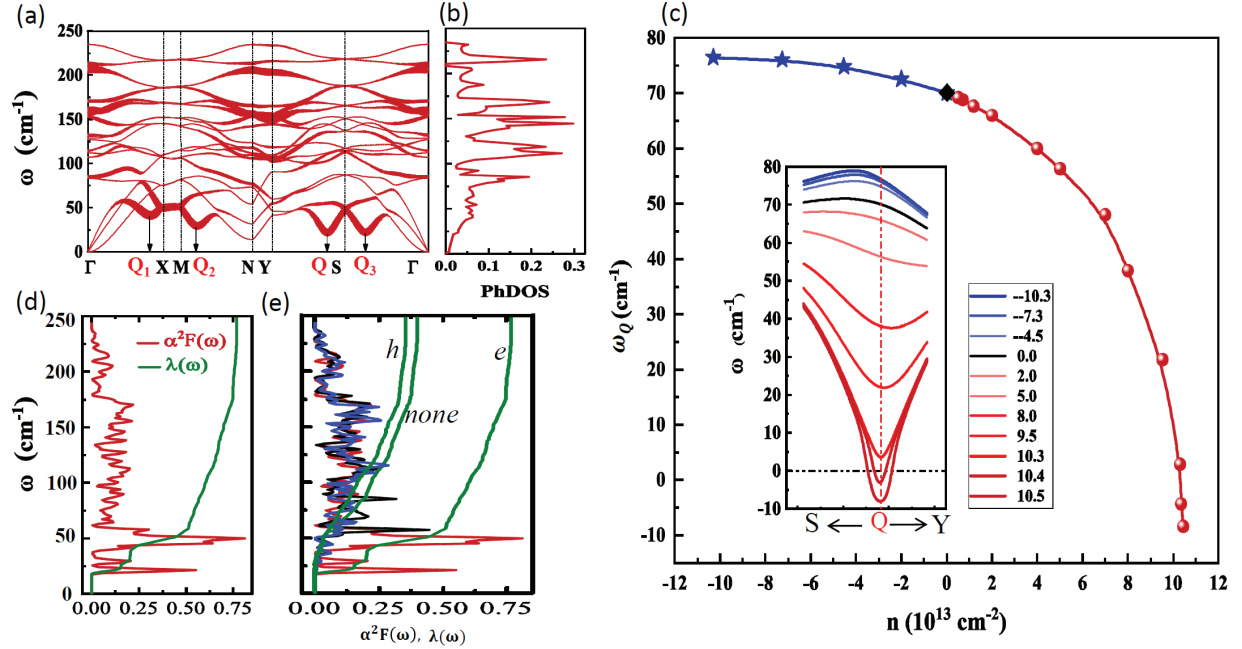}
\end{center}
\caption{(Color online) (a) The calculated phonon spectra for $e$-doped
monolayer 1T$^{\prime}$-WTe$_{2}$ with $n_{e}=9.5\times10^{13}$\textit{cm}%
$^{-2}$. The linewidth of the spectrum $\gamma_{q,\nu}$ is indicated by the
weight of the curves. All $Q$ and $Q_{i}$ phonon modes have $(0.73\pi
/a,q_{y})$. (b) The phonon density of states of spectrum in (a).(c) The
frequency $\omega_{Q}$ of $Q$ mode as a function of carrier concentration of
electrons (red ball) and holes (blue star). The inset depicts the evolution of
the $Q$ mode with different charge densities. (d)The calculated $\alpha
^{2}F(\omega)$ and $\lambda(\omega)$ with the same electron-doped density in
(a). (e) The calculated $\alpha^{2}F(\omega)$ with blue, dark and red for
hole-, non-, and electron-doped cases, respectively and $\lambda(\omega)$ for
hole- (h), none-, and electron- (e) doping for comparison. The electron- and
hole-doped densities correspond to those in Fig.\ref{f1} (c) and (d),
respectively.}%
\label{f2}%
\end{figure}

The calculated phonon spectrum $\omega_{q,\nu}$ with wave vector $q$ and mode
index $\nu$ in electron-doped 1T$^{\prime}$-WTe$_{2}$ is shown in Fig.
\ref{f2}(a). Remarkably, some phonon modes at specific wave vectors $Q$ and
$Q_{i}$ with $i$=1,2,3 dramatically soften in comparison with the non-doped
case shown in Fig. S4 (a) in SM. However, the phonon spectrum of hole doping
slightly stiffens, as shown in Fig. S4(d) in SM. This distinctly opposite
trend of the phonon modes between electron doping and hole doping is clearly
demonstrated in Fig. \ref{f2}(c). The softening of $Q$ mode phonon is enhanced
as electron doping density increases, and its frequency $\omega_{Q}$ turns to
be negative when $n_{e}$ crosses $10.3\times10^{13}$ cm$^{-2}$, as shown in
the inset in Fig. \ref{f2}(c). It means some long-range order instability
could emerge. In the present case, such instability is probably the CDW.
Furthermore, the relevant atomic displacements for the $Q$ mode phonon are
mainly expressed as horizontal W-W stretching vibration and shear Te-Te
tortuosing vibration, as shown in Fig. S5 in SM. These behaviors indicate the
$Q$ mode phonon has the longitudinal-wave feature. We have checked that the
properties of all $Q_{i}$ mode phonon are similar to those of the $Q$ mode phonon.

\begin{figure}[ptb]
\begin{center}
\includegraphics[width=1\columnwidth]{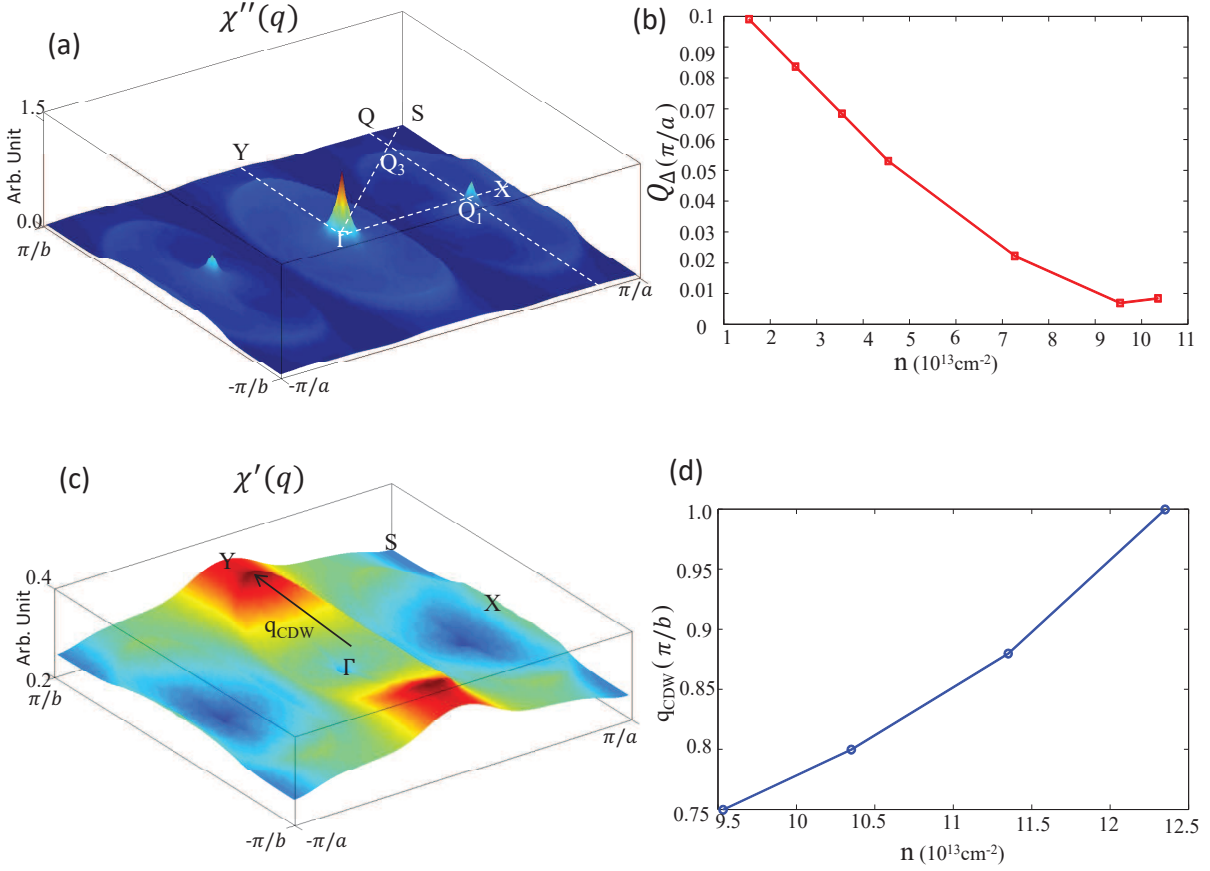}
\end{center}
\caption{(Color online)(a) and (c) The imaginary and real electron
susceptibility $\chi^{\prime\prime}(q)$ and $\chi^{\prime}(q)$ for the
electron doping with $n_{e}=9.5\times10^{13}$\textit{cm}, respectively. Note
that the peak of $\chi^{\prime\prime}(q=0)$ in (a) is trivial. (b) The
evolution of $Q_{\Delta}=|q_{nest,x}-0.73\pi/a|$ versus the electron-doped
level. (d) The evolution of CDW wave vector versus the electron-doped level.}%
\label{f3}%
\end{figure}

\begin{figure}[ptb]
\begin{center}
\includegraphics[width=1\columnwidth]{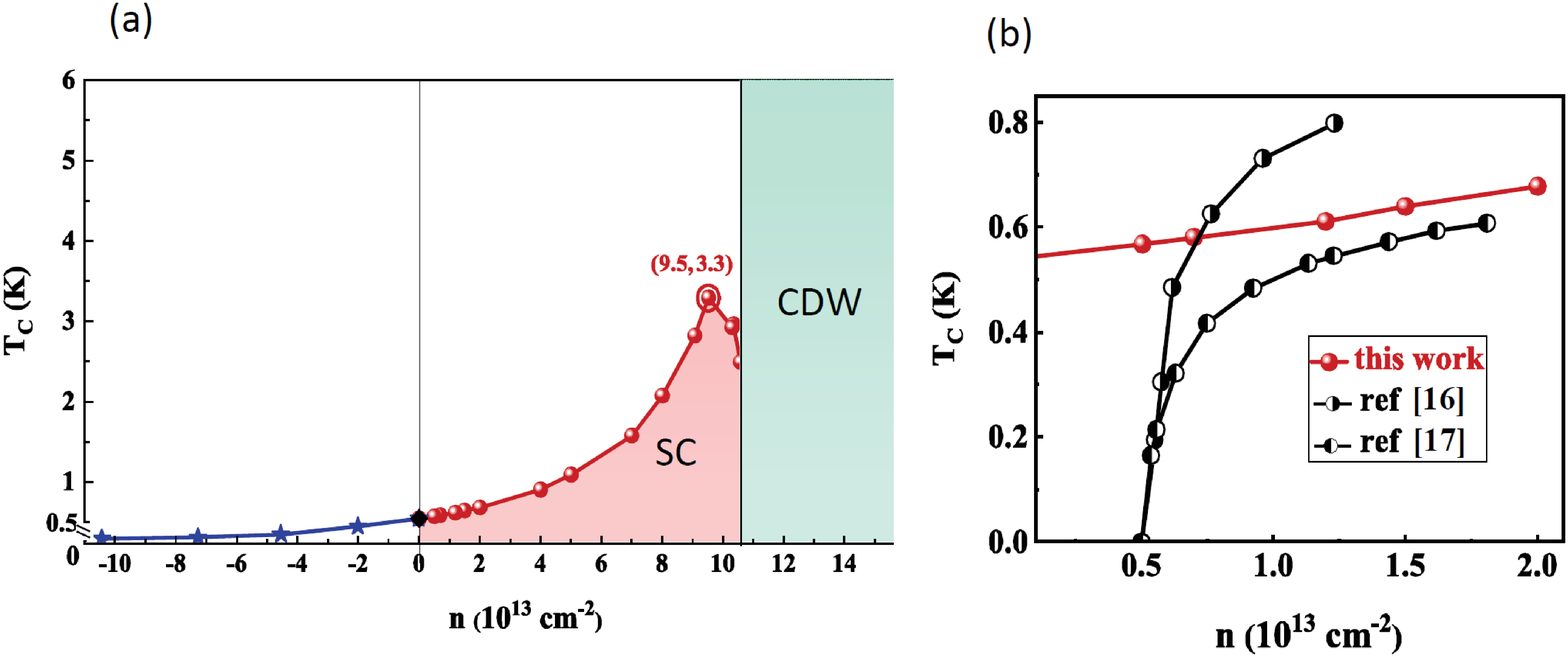}
\end{center}
\caption{(Color online) (a) A $T_{c}-n$ phase diagram is plotted. The boundary
of solid line with small balls are given by our calculations. SC and CDW label
superconducting phase and charge density wave state, respectively. (b) A
comparison between our calculation with the experimental results
\cite{Sajadi-s-2018, fatemi-s-2018} for $e$-doped 1T$^{\prime}$-WTe$_{2}$. The
logarithmical $T_{c}$ from \cite{Sajadi-s-2018} has been converted.}%
\label{f4}%
\end{figure}

The features of phonon spectrum in Fig. \ref{f2} imply that the soft mode
phonons should play a key role to induce the superconductivity in
electron-doped monolayer 1T$^{\prime}$-WTe$_{2}$. To prove it, we also plot
the linewidth of the phonon spectrum $\gamma_{q,\nu}$ shown in Fig.
\ref{f2}(a), the Eliashberg spectral function $\alpha^{2}F(\omega)$ and the
frequency-dependent electron-phonon coupling (EPC) strength $\lambda(\omega)$
shown in Fig. \ref{f2}(d) and (e). The details about $\gamma_{q,\nu}$,
$\alpha^{2}F(\omega)$\cite{allen-prb-1972}, and $\lambda(\omega)$ are shown in
Section III in SM. With them, one can get a crucial expression for the
effective attractive interaction as follows,
\begin{equation}
V_{q,\nu}^{ep}=-4\gamma_{q,\nu}/(\pi N_{0}^{2}\omega_{q,\nu}^{2}).
\label{ep interaction}%
\end{equation}
Here, $N_{0}$ is the electronic density of states at the Fermi surface for
both spin orientations. From Fig. \ref{f2}(a), and Fig. S4 (a) and (d) in SM,
one can find that some optical modes have large linewidth $\gamma_{q,\nu}$ for
all three cases. The relevant frequency $\omega_{q,\nu}$, however, is very
high. From Eq. (\ref{ep interaction}), the $V_{q,\nu}^{ep}$ from the optical
modes can be neglected. The dramatic difference occurs for the longitudinal
acoustic mode. One can find that the electron doping can greatly enlarge the
linewidth $\gamma_{q,\nu}$ and soften the frequency $\omega_{q,\nu}$ at the
specific momenta $Q$ and $Q_{i}$. Therefore, the $\alpha^{2}F(\omega)$ and
$\lambda(\omega)$ are strongly enhanced by electron doping, as shown in Figs.
\ref{f2}(d), 2(e) and Fig. S4(c), (f) in SM, and so is $V_{q,\nu}^{ep}$. In
terms of superconductivity, only the electrons very close to the Fermi surface
need to be considered. Such constraint demands the good nesting condition for
the electron-doped Fermi surface in Fig. \ref{f1} (d), which can be clarified
by the electron susceptibility $\chi(q)=\chi^{\prime}(q)+i\chi^{\prime\prime
}(q)$ (See Section V in SM for details). The Fermi surface nesting is
evaluated by the imaginary part $\chi(q)$, i.e., $\chi^{\prime\prime}(q)$. In
Fig. \ref{f3} (a), we plot the pattern of $\chi^{\prime\prime}(q)$ for
electron-doped case with $n_{e}=9.5\times10^{13}$cm$^{-2}$. The nesting wave
vector $q_{nest}=(0.73\pi/a,0)$, which is quite close to horizontal ordinate
$\sim0.73\pi/a$ of all $Q$ and $Q_{i}$ mode phonons. We also plot the
evolution of $Q_{\Delta}=|q_{nest,x}-0.73\pi/a|$ as the doping density in Fig.
\ref{f3} (b) and Fig. S6 in SM. One can find that the best nesting with
$Q_{\Delta}\sim0$ corresponds to the optimal superconductivity with the
highest transition temperature $T_{c}$, as shown in Fig. \ref{f4} (a). The
consistent relationship between the phonon spectrum and the electronic
structure indicates that the Fermi surface nesting plays the important role to
drive the superconductivity in 1T$^{\prime}$-WTe$_{2}$.

Furthermore, we calculate the superconducting transition temperature $T_{c}$
in different doping levels using the McMillan-Allen-Dynes formula
\cite{allen-prb-1975}. The details are presented in Section VI in SM. The
calculated results (data points) of $T_{c}$ as a function of density of
electrons (red ball) and holes (blue star) are plotted in Fig. \ref{f4}(a).
Clearly, compared to the hole doping, the electron doping indeed strengthens
the $T_{c}$ of monolayer 1T$^{\prime}$-WTe$_{2}$, showing significant
asymmetry features. More interestingly, the $T_{c}$ curve shows two remarkable
features. First, the $T_{c}$ exponentially increases as pumping electrons. The
nonlinear behavior of $T_{c}$ curve is further supported by the exponential
decay trend of logarithmic average frequency $\omega_{\log}$ and the
exponential ascend trend of the EPC constant $\lambda^{ph}$, as shown in Fig.
S8 in SM. The behavior of $T_{c}$ curve provides the possibility to obtain the
quite high $T_{c}$ superconductivity through increasing electron doping, which
is different from the doped graphene \cite{si-prl-2013} or doped antimonene
\cite{lugovskoi-prb-2019}. Second, $T_{c}$ reaches the maximum of 3.3 K with
$n_{e}=9.5\times10^{13}$ cm$^{-2}$ corresponding to the best nesting
condition, then drops, and finally disappears or merges into the possible CDW
state. Here, we discuss a little more about the CDW instability. Recall that
the frequency $\omega_{q}$ of $Q$ mode phonon becomes negative when $n_{e}$
increases cross $10.3\times10^{13}$ cm$^{-2}$, as shown in Fig. \ref{f1}(c).
It indicates the lattices become unstable and the system undergoes a possible
CDW transition. For the electronic states, the CDW instability is induced by
the divergence of $\chi^{\prime}(q)$, the real part of electron susceptibility
$\chi(q)$\cite{Johannes-prb-2008}. In Fig. \ref{f3} (c), we plot the pattern
of $\chi^{\prime}(q)$ for electron-doped case with $n_{e}=9.5\times10^{13}%
$\textit{cm}$^{-2}$, whose CDW wave vector is $q_{CDW}=(0,0.75\pi/b)$. When
the electron doping increases, $q_{CDW}$ is more and more close to the
$(0,\pi/b)$, as shown in Fig. \ref{f3}(d) and Fig. S7 in SM. We conclude that
the CDW with different $q_{CDW}$ should compete with each other, and the most
probable one has the wave vector $(0,\pi/b)$, which is quasi-one dimensional,
commensurate $2Q$ CDW. Note that the nesting wave vector has nothing to do
with the CDW wave vector\cite{Johannes-prb-2008}. According to the above
discussions, the relevant phase diagram can be plotted as shown in Fig.
\ref{f4} (a), from which a superconducting dome can be found. Currently, using
solid-state gates or ionic-liquid gates, typical carrier density ($10^{14}$
cm$^{-2}$) can be achieved in several two-dimensional (2D) materials
\cite{Fu-npj-2017,Li-n-2016,Saito-st-2016,Lu-s-2015,Costanzo-nn-2016}, we thus
expect that future experiments for electron-doped 1T$^{\prime}$-WTe$_{2}$ can
verify our predicted maximum $T_{c}$ (3.3 K), which is about four or five
times the experimental values of 0.8 K \cite{Sajadi-s-2018} or 0.6 K
\cite{fatemi-s-2018}. In the range of experimental charge density, i.e.,
$n\in\lbrack0.7,2.1]\times10^{13}$cm$^{-2}$, our results are in qualitatively
agreement with the experimental values \cite{Sajadi-s-2018,fatemi-s-2018}, as
shown in Fig. \ref{f4} (b). Whereas $T_{c}$ greatly deviates from the
experimental value below $0.7\times10^{13}$cm$^{-2}$. It is because the
Coulomb screening effect is strongly reduced in the low density regime.
Correspondingly, the constant retarded Coulomb pseudo-potential $\mu^{\ast}$
is not exact and should become large to reduce $T_{c}$ from Eq.(7) in SM.

Now, we turn to understand the superconducting properties of monolayer
1T$^{\prime}$-WTe$_{2}$ in the electron-doped regime. According to Fig.
\ref{f1}(d), there exist two separated Fermi pockets centered at momenta
$\mathbf{K}^{\prime}$ and $\mathbf{K}$ which can be regarded as two valleys.
The low-energy Hamiltonian can be expanded around $\mathbf{K}^{\prime}$ and
$\mathbf{K}$. Define the annihilation electron operator as $c_{\mathbf{k}%
,\eta,s}$, where $\mathbf{k}$ is the momentum measured from $\mathbf{K}%
^{\prime}$ and $\mathbf{K}$, $\eta=\pm$ is for two valleys and $s=\uparrow
$,$\downarrow$ is for spin. On the basis of $[c_{\mathbf{k},+,\uparrow
},c_{\mathbf{k},-,\uparrow},c_{\mathbf{k},+,\downarrow},c_{\mathbf{k}%
,-,\downarrow}]$, the effective Hamiltonian is%
\begin{equation}
\mathcal{H}_{0}=(\varepsilon_{k}-\mu)s_{0}\tau_{0}+g\mu_{B}\tau_{0}%
\mathbf{B}\cdot\mathbf{s}. \label{H0}%
\end{equation}
Here, $s$ and $\tau$ are two sets of Pauli matrices for spin and valley
degrees of freedom, $\varepsilon_{k}=\hslash^{2}k^{2}/2m-\mu$ is the
dispersion with the chemical potential $\mu$, $g$ and $\mu_{B}$ are the Lande
$g$ factor and Bohr magneton, respectively, and $\mathbf{B}=(B_{x},B_{y}%
,B_{z})$ is the external magnetic field. Note that we only take into account
the Zeeman effect of the external magnetic field, but neglect the orbital
effect due to the 2D feature of 1T$^{\prime}$-WTe$_{2}$. In the
superconducting states, the pairings should follow the IRs of the $C_{2h}$
point group of 1T$^{\prime}$-WTe$_{2}$. In order to consider the
superconductivity for simplicity, we adopt the approximation of Fermi surface
average of $V_{q,\nu}^{ep}$ to neglect the weak momentum dependence,
\textit{i.e.}, $V^{ep}=\sum_{\nu}$ $\langle V_{\mathbf{k}-\mathbf{k}^{\prime
},\nu}^{ep}\rangle_{\mathbf{k},\mathbf{k}^{\prime}\in FS}$. Thus, we only
consider the momentum-independent $s$-wave pairings, which are from the
constant pairing interaction $-V_{0}=V^{ep}+U_{scou}$ with $U_{scou}$ the
screened Coulomb interaction. Furthermore, the soft-mode phonons have the
momenta connecting the two valleys, which indicates only the inter-valley
pairing is possible. Under the mean-field approximation in the Nambu basis, we
can classify the pairing symmetry under the constraint of anti-commutation
relation between fermion operators. The pairing function can be parameterized
by the form $\hat{\Delta}=\sum_{i}\Delta_{i}\Gamma_{i}$, where $\Gamma_{i}$ is
the $i$th IR matrix product from $s$ and $\tau$. The results are listed in
Table \ref{tb1}. More details are shown in Section VII in SM.
\begin{table}[ptb]
\caption{Classification of the inter-valley $s$-wave pairings according to the
representations of $C_{2h}$ point group.}%
\label{tb1}%
\begin{tabular}
[c]{llllll}\hline\hline
& $C_{2}(z)$ & $\sigma_{h}$ & $i$ & $\ \ \ \ \ \hat{\Delta}$ & $\ \ \ \ \ \hat
{\Delta}$\\\hline
IR & $s_{z}\tau_{x}$ & $-is_{z}$ & $\tau_{x}$ & Matrix form & Explicit form\\
$A_{g}$ & 1 & 1 & 1 & $\ \ \ \ \ \tau_{x}$ & $c_{+\uparrow}c_{-\downarrow
}-c_{+\downarrow}c_{-\uparrow}$\\
$B_{u}$ & -1 & 1 & -1 & $\ \ \ \ s_{z}\tau_{y}$ & $c_{+\uparrow}%
c_{-\downarrow}+c_{+\downarrow}c_{-\uparrow}$\\
$A_{u}$ & 1 & -1 & -1 & $\ \ \ \ s_{x}\tau_{y}$ & $c_{+\uparrow}c_{-\uparrow
}-c_{+\downarrow}c_{-\downarrow}$\\
&  &  &  & $\ \ \ \ s_{y}\tau_{y}$ & $i(c_{+\uparrow}c_{-\uparrow
}+c_{+\downarrow}c_{-\downarrow})$\\\hline\hline
\end{tabular}
\end{table}The pairing interaction in all pairing channels in Table \ref{tb1}
has the same amplitude of $V_{0}/2$.

To evaluate the possible superconducting pairing in 1T$^{\prime}$-WTe$_{2}$,
we solve the following linearized gap equations for $T_{c}$%
\cite{Fu-prl-2010,Liu-prl-2017},%

\begin{equation}
V_{0}\chi_{i}/2=1, \label{gf1}%
\end{equation}
where $\chi_{i}$ is the finite temperature superconducting susceptibility in
$i$th pairing channel, and can be calculated by%

\begin{equation}
\chi_{i}=-\frac{1}{\beta}\sum_{i\omega_{n},k}Tr[\Gamma_{i}^{\dag}%
\mathcal{G}_{e}(k,i\omega_{n})\Gamma_{i}\mathcal{G}_{h}(k,i\omega_{n})].
\label{s1}%
\end{equation}
Here, $\mathcal{G}_{e/h}(k,i\omega_{n})=[i\omega_{n}\mp\mathcal{H}%
_{0}(k)]^{-1}$ are the relevant standard electron and hole Matsubara Green
functions. The details are shown in Section VII in SM, and we only list the
key results here. Note that only the electron-type bands are taken into
account in the electron-doped regime. With the approximation that the pairing
occurs at the Fermi surface, we can get $\chi_{A_{g}}=4\chi_{0}$, $\chi
_{B_{u}}=4\chi_{0}B_{z}^{2}/B^{2}$, $\chi_{A_{u}}=4\chi_{0}B_{x}^{2}/B^{2}$ or
$4\chi_{0}B_{y}^{2}/B^{2}$. $\chi_{0}=N_{0}\int d\varepsilon\tanh
(\frac{\varepsilon}{2k_{B}T})/\varepsilon$ is the standard superconducting
susceptibility and $k_{B}$ is the Boltzmann constant. When $\mathbf{B}=0$, one
can find that all the pairing channels have the same superconducting
susceptibility, i.e., $4\chi_{0}$, likewise, the same $T_{c}$, i.e.,
$k_{B}T_{c}=\frac{2e^{\gamma}}{\pi}\hbar\omega_{D}\exp(-\frac{1}{2N_{0}V_{0}%
})$, where $\gamma\approx0.5772$ is the Euler constant and $\omega_{D}$ is the
Debye frequency. It indicates that all the pairing channels are degenerate and
are the possible candidate for the superconducting ground state. When
$\mathbf{B}\neq0$, the $T_{c}$ of $A_{g}$ channel is robust against the
magnetic field. The $T_{c}$ of $B_{u}$ or $A_{u}$ channel is
magnetic-field-direction dependent. Namely, $\ln\frac{T_{c,B_{u}}%
(B)}{T_{c,B_{u}}(0)}=\frac{1}{2N_{0}V_{0}}(1-\frac{B^{2}}{B_{z}^{2}})$ and
$\ln\frac{T_{c,A_{u}}(B)}{T_{c,A_{u}}(0)}=\frac{1}{2N_{0}V_{0}}(1-\frac{B^{2}%
}{B_{x/y}^{2}})$. Therefore, the $B_{u}$ channel is robust against the
out-of-plane magnetic field $B=(0,0,B_{z})$ and is fragile for the in-plane
magnetic field $B=(B_{x},B_{y},0)$. In other words, $B_{u}$ channel has large
$H_{c2,\perp}$ and small $H_{c2,\Vert}$. Similarly, one can find $A_{u}$
channel has small $H_{c2,\perp}$ and large $H_{c2,\Vert}$. In comparison with
the experimental observations, it inidicates the superconducting state should
fall into the $A_{u}$ channel with the equal-spin-triplet pairing, whose
$H_{c2}$ cannot be restricted by the Pauli paramagnetic limit. These results
can be easily understood through the spin structure of the Cooper pairs. The
$B_{u}$ and $A_{u}$ channels correspond to the three components $S_{z}$ and
$S_{x}$, $S_{y}$ of the $S=1$ spin-triplet Cooper pairs, respectively.
Therefore, the couplings between them and the magnetic field follow the
Zeeman-type to minimize the ground state energy. For a general magnetic field
with $B=B_{0}(\cos\theta\cos\varphi,\cos\theta\sin\varphi,\sin\theta)$, we can
get the exact pairing form as $\hat{\Delta}\sim e^{i\varphi}c_{\mathbf{k}%
,+\uparrow}c_{-\mathbf{k},-\uparrow}-e^{-i\varphi}c_{\mathbf{k},+\downarrow
}c_{-\mathbf{k},-\downarrow}$, where the phase $\varphi$ is determined by the
direction of magnetic field.

There exist several other mechanisms to result in the anisotropic $H_{c2}$
including type-I \cite{Lu-s-2015,Saito-np-2016,Xi-np-2016}, type-II Ising
pairings \cite{Falson-s-2020} and spin-orbit scattering \cite{Klemm-prb-1962}.
Type-I Ising pairing mechanism is related to inversion symmetry breaking and
clearly is not the case in monolayer 1T$^{\prime}$-WTe$_{2}$, which preserves
the inversion symmetry. Type-II Ising pairing mechanism relies on the band
near $\Gamma$ point with out-of-plane orientation of the spin locked by the
spin-orbit coupling. This mechanism can also be excluded, because no such kind
of band splitting exists in monolayer 1T$^{\prime}$-WTe$_{2}$. For the
spin-orbit scattering mechanism, there are two kinds of situations. In the
clean limit of crystalline sample with high mobility, the spin-orbit coupling
scattering can be discarded as it could induce the unphysically short
scattering times \cite{Lu-s-2015,Saito-np-2016,Xi-np-2016}. In the dirty
limit, the strong spin-orbit coupling indeed could induce the anisotropic
$H_{c2}$. However, the superconductivity refers to the Fermi surface connected
by the momenta relating to soft phonons. From Fig. \ref{f1} (a), the pieces of
the Fermi surfaces have strong effective spin-orbit coupling is close to
$\Gamma$ point, whereas the pieces of the Fermi surfaces connected by nesting
wave vector has weak effective spin-orbit coupling \cite{Qian-s-2014}. This is
also the reason why the simple low-energy Hamiltonian in Eq. (\ref{H0}) is
enough and is adopted to investigate the properties of superconducting state
in monolayer 1T$^{\prime}$-WTe$_{2}$. Thus, the spin-orbit scattering
mechanism can be excluded.

In conclusion, we have determined the soft-mode phonons, the relevant enhanced
electron-phonon coupling and the Fermi surface nesting are three ingredients
to drive the superconductivity in electron-doped monolayer 1T$^{\prime}%
$-WTe$_{2}$. We also predict a quasi-one dimensional commensurate $2Q$ CDW
emerges in the heavy electron doping regime, and a superconducting dome in the
phase diagram. Furthermore, we propose an unconventional superconducting
pairing with equal spin triplet to capture the experimentally observed strong
anisotropic upper critical fields. Our studies provide a standard way to
understand the gating-dependent superconductivity in 1T$^{\prime}$-WTe$_{2}$
and other transition metal dichalcogenides.

\begin{acknowledgments}
This work was financially supported by the National Key R\&D Program of China
No. 2017YFA0303201, National Natural Science Foundation of China under Grants
(No. 11625415, No. 61605014, No. 11674331), the \textquotedblleft Strategic
Priority Research Program (B)\textquotedblright\ of the Chinese Academy of
Sciences, Grant No. XDB33030100, the `100 Talents Project' of the Chinese
Academy of Sciences, the Collaborative Innovation Program of Hefei Science
Center, CAS (Grants No. 2020HSC-CIP002), the CASHIPS Director's Fund
(BJPY2019B03), the Science Challenge Project under Grant No. TZ2016001. A
portion of this work was supported by the High Magnetic Field Laboratory of
Anhui Province, China.
\end{acknowledgments}

\end{document}